\documentclass[12pt]{article}
\usepackage{sao1}
\usepackage{psfig}

\usepackage{amsmath}

\newcommand{\zav}[1]{\left(#1\right)}

\DeclareMathAlphabet{\mathsc}{OT1}{cmr}{m}{sc}
\DeclareRobustCommand{\ion}[2]{%
\relax\ifmmode
{\mathbf{#1\,\mathsc{#2}}}
\else\textup{#1\,{\sc{#2}}}%
\fi}

\newcommand\hvezda{HD~37776}

\begin{document}

\title{The light variations of \hvezda~as a result of the uneven surface
distribution of helium and silicon}

\author{Krti\v{c}ka J.\inst{1} \and Mikul\'{a}\v{s}ek Z.\inst{1} \and
Zverko J.\inst{2} \and \v{Z}i\v{z}\v{n}ovsk\'{y}\inst{2}}

\institute{
Institute of Theoretical Physics and Astrophysics,
    Kotl\'{a}\v{r}sk\'{a} 2, CZ-637 00 Brno, Czech Republic
\and
Astronomical Institute, Slovak Academy of Sciences,
    SK-059 60 Tatransk\'{a} Lomnica, Slovak Republic
}

\maketitle

\begin{abstract}

We calculate the $uvby$ light curves of the helium-strong chemically peculiar
(CP) star \hvezda. Assuming that the chemical peculiarity influences the
monochromatic radiative flux due to mainly the bound-free processes and using
the model of the surface distribution of helium and silicon on \hvezda, as
derived from spectroscopy, we calculate the emergent fluxes for a series of its
rotational phases. We show that it is possible to consistently simulate light
curves of a chemically peculiar star and that the basic properties of
variability of helium CP stars can be understood in terms of the model of spots
with a peculiar chemical composition.

\end{abstract}

\section{Introduction}

The periodic light variations are a common feature among the magnetic CP stars.
It is believed that most of these variations are caused by the rotational
modulation of the observed radiative flux due to spots of a different peculiar
chemical composition present on the stellar surface. In \cite{mypoprad} we
investigated the effect of the uneven surface distribution of helium on the
monochromatic emergent flux and showed that 
the distribution of the flux may be changed by
the bound-free processes. In this work we calculate the light curves of
helium-strong CP star
\hvezda\
using the spectroscopically derived surface distribution
of
helium and
silicon by \cite{choch}. 
To study the influence of the
chemical peculiarity on the observed light variability besides helium we 
selected silicon as:
(i) it is overabundant for a large group of chemically peculiar stars that show
light variations (e.g., Si, He-weak, and He-strong stars), and these stars form
a photometrically homogeneous group; (ii) a large overabundance of silicon is
observed on the \hvezda\ surface.


\section{Atmosphere models of \hvezda}

Given the map of the helium and silicon surface distribution we computed the
emergent flux for the every surface element, sized $1^\circ\times1^\circ$, and
then integrated over the all visible surface. The fluxes were computed using
SYNSPEC, the code for spectrum synthesis, for $8\times10$ LTE models computed
with TLUSTY (\cite{hublad}) for different silicon and helium abundance from the
grid $\text{[Si/H]}=-3, -2, -1, 0, 1, 2, 2.5, 3$ and $\text{[He/H]}=-5, 
-4, -3, -2,
-1, 0, 1, 2, 2.5, 3$~ to cover the extent of values given by \cite{choch}. The
effective temperature ${{T}_\mathrm{eff}}$ and surface gravity ${\log g}$ of the
models are the same (see Tab.~\ref{hvezpar}). The atomic species and the number
of their energy levels included in the model atmosphere calculations are listed
in Tab.~\ref{ionpar}.

\begin{table}[hbt]
\caption{Parameters of \hvezda}
\label{hvezpar}
\begin{center}
\begin{tabular}{lccc}
\hline
${{T}_\mathrm{eff}}$
& ${22\,000}$\,K & \cite{groka}\\
${\log g}$
& ${4.0}$ & \cite{groka}\\
inclination ${i}$ & ${45^\circ}$ &
\cite{choch}\\
spots parameters  & &\cite{choch}\\
\hline
\end{tabular}
\end{center}
\end{table}

\begin{table}[hbt]
\caption{The atomic species and the number of their energy levels included in
the model atmosphere calculations}
\label{ionpar}
\begin{center}
\begin{tabular}{lclclclc}
\hline\hline
Ion & Levels & Ion & Levels  & Ion & Levels & Ion & Levels  \\
\hline
\ion{H}{i}   &   9&\ion{N}{i}   & 21 &\ion{Ne}{i}  & 15 &\ion{Si}{ii}  & 40 \\
\ion{H}{ii}  &  1 &\ion{N}{ii}  & 26 &\ion{Ne}{ii} & 15 &\ion{Si}{iii} & 30 \\
\ion{He}{i}  & 24 &\ion{N}{iii} & 32 &\ion{Ne}{iii}& 14 &\ion{Si}{iv}  & 23 \\
\ion{He}{ii} & 20 &\ion{N}{iv}  &   1&\ion{Ne}{iv} &   1&\ion{Si}{v}   & 1\\
\ion{He}{iii}&   1&\ion{O}{i}   & 12 &\ion{Mg}{i}  & 13 &\ion{S}{ii}  & 14 \\
\ion{C}{i}   & 26 &\ion{O}{ii}  & 13 &\ion{Mg}{ii} & 14 &\ion{S}{iii} & 10 \\
\ion{C}{ii}  & 14 &\ion{O}{iii} & 29 &\ion{Mg}{iii}& 14 &\ion{S}{iv}  & 15 \\
\ion{C}{iii} & 12 &\ion{O}{iv}  &   1&\ion{Mg}{iv} &   1&\ion{S}{v}   &   1\\
\ion{C}{iv} &     1 \\
\hline
\end{tabular}
\end{center}
\end{table}

We calculate the flux from model atmosphere for given silicon and helium
abundances $H(\lambda,\text{[He/H],[Si/H]})$. The flux
$H_c(\text{[He/H],[Si/H]})$ in the color $c$ is obtained as a convolution of
$H(\lambda,\text{[He/H],[Si/H]})$ with a Gauss profile peaked at the central
wavelength of the color ${\lambda_c}$, with dispersion ${\sigma\!}_c$. The
radiative flux in the color ${c}$ from individual surface elements
${H_c(\Omega)}$ ($\Omega$ are spherical coordinates on the stellar surface) is
obtained by interpolation of fluxes from the model grid. Finally, the total
observed radiative flux is calculated as the integral over all the visible
surface elements
\begin{equation}
{H_c=\int H_c(\Omega)\,u(\theta)\cos\,\theta\,
\mathrm{d}\Omega}
\end{equation}
($\theta$ is the angle between normal to the surface element
and line of sight, and $u(\theta)$ describes the adopted linear limb darkening) and the observed
magnitude difference is then
\begin{equation}
{\Delta
m_{c}=-2.5\,\log\,\zav{\frac{{H_c}}{H_c^\mathrm{ref}}},}
\end{equation}
where ${H_c^\mathrm{ref}}$ is a reference flux.
This was performed for $36$ rotational phases (some of them are shown on
Fig.~\ref{povrch}).

\section{The origin of light variations}

\begin{figure}[t]
\centerline{\psfig{figure=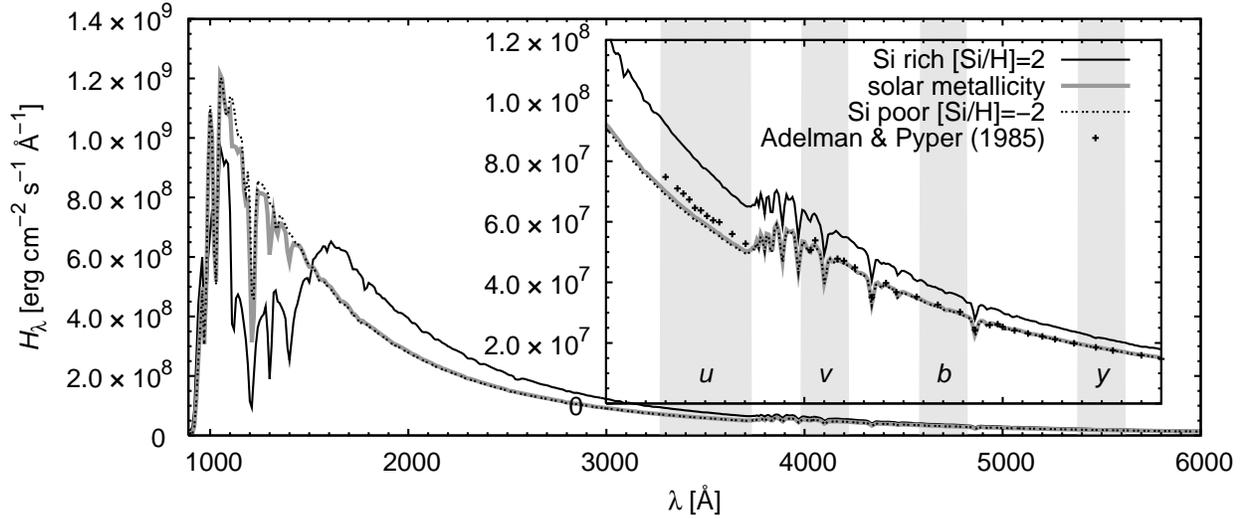,height=7.cm}}
\caption{The radiative fluxes from the \hvezda\ atmosphere calculated with 
different silicon abundances and the observed averaged dereddened flux 
distribution according to \protect\cite {adelpy}. The radiative flux was 
smoothed to better demonstrate
the light separation. In the {\it uvby} colors, the silicon spots are 
evidently brighter than the regions with low silicon abundance. 
Silicon abundances lower than the solar do not significantly change 
the radiative flux.}
\label{toky}
\end{figure}

The plot of the radiative flux calculated for  different values of the silicon
abundance (Fig.~\ref{toky}) shows that stellar surface in silicon rich regions
is brighter in the {\it uvby} colors than in the silicon poor ones. The increase of
the flux in the optical domain is caused by the flux redistribution from the far
UV region
mainly
due to the enhanced silicon bound-free absorption.

\section{Elemental surface distribution}

\begin{figure}[t]
\centerline{\psfig{figure=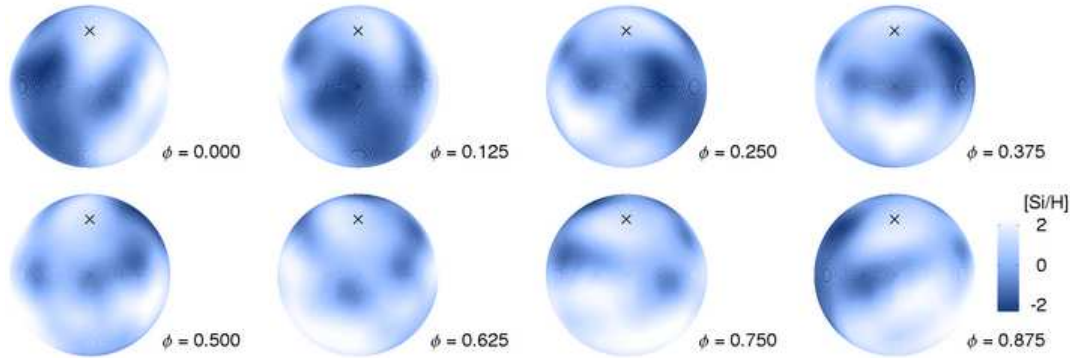,height=5.cm}}
\caption{Variation of the silicon abundance on the surface of \hvezda\ derived
from spectroscopy by \cite[smoothed]{choch} for different rotational phases.
The Si overabundant regions are bright on the stellar surface (see
Fig.~\ref{toky}) and white on the graph; the Si underabundant regions are dark
on the stellar surface and blue (dark) on the graph.}
\label{povrch}
\end{figure} 

At different rotational phases, surface elements with different silicon
abundance are seen on the stellar disc. During the light maximum around the
phase $\phi=0.75$  (see also Fig.~\ref{hvvel}) the Si rich regions are seen,
which have large flux in the {\it uvby} (see Fig.~\ref{toky}). During the
light minimum around the phase $\phi=0.1$ the Si poor regions are seen on the
stellar disc, which have lower flux in the {\it uvby}. 

\section{Predicted light variations}

Due to the uneven surface distribution of silicon (and partly also due to
helium) the emergent monochromatic flux varies with the location on the 
visible disc. As shown in Fig.~\ref{toky}, the Si rich regions are brighter 
in the {\it uvby} colors (the brightness temperature is higher), whereas Si 
poor regions are darker.

\begin{figure}[t]
\centerline{\psfig{figure=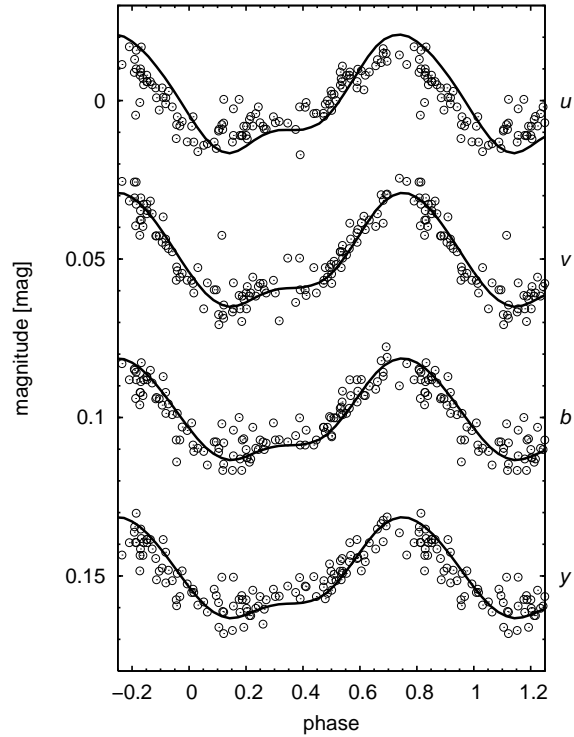,height=10.cm}}
\caption{The predicted light variations of \hvezda\ calculated using the 
silicon
and helium surface distribution after \cite{choch} and compared with the 
observed ones (empty circles).}
\label{hvvel}
\end{figure}

Using the model atmosphere fluxes derived for the different helium and silicon
abundances we calculated the theoretical light curves (see Fig.~\ref{hvvel}) and
compared them with the observed ones. The observed light variations can be
explained as alternation of the emergent flux due to the uneven surface
distribution of silicon and helium.

\section{Conclusions}

We are able to simulate the light variations of the He-strong CP star \hvezda\
assuming that the light variations are due to the uneven surface distribution of
silicon and helium. The silicon and helium spots modify the emergent flux mainly
due to bound-free transitions. The predicted light curves reproduce the observed
ones very well in their overall shape and amplitude. We stress that the
theoretical light curves were obtained from the spectroscopical silicon surface
distribution without using any free parameter. We were able for the first time
(to our knowledge) to simulate the light curve of a CP star. Due to the
satisfactorily good agreement between the observed and predicted light
variations we conclude that this modelling is a very promising way towards the
explanation of the light variations of chemically peculiar stars.


\begin{acknowledgements}
This work was supported by grants 
GA\v{C}R 205/06/0217 and VEGA 2/6036/6.
This research has made use
of NASA's Astrophysics Data System, the SIMBAD database, operated at
CDS, Strasbourg, France and \emph{On-line database of photometric
observations of mCP stars} (\cite{mikdat}).
\end{acknowledgements}

\end{document}